\documentclass[runningheads]{llncs}

\usepackage[T1]{fontenc}
\usepackage{graphicx}
\usepackage{booktabs}
\usepackage{amsmath}
\usepackage{amssymb}
\usepackage{hyperref}
\usepackage{xcolor}
\usepackage{tikz}
\usetikzlibrary{positioning,arrows.meta,shapes.geometric,fit,calc,decorations.pathreplacing}
\usepackage{CJKutf8} % For Korean characters in Case B

\begin{document}

\title{SLIP \& ETHICS: Graduated Intervention\texorpdfstring{\\}{ }for AI Emotional Companions}

\author{Minseo Kim}
\authorrunning{M. Kim}
\institute{HUA Labs, Seoul, Republic of Korea\\
\email{devin@hua-labs.com}}

\maketitle

% --- Abstract ---

\begin{abstract}
AI emotional companions face a fundamental safety-rapport paradox: overly restrictive safety measures compromise therapeutic alliance, while permissive systems risk user harm. We provide preliminary evidence that structured qualitative indicators---affect intensity ($a$), narrative dynamism ($m$), and contextual signal tags---can help grade the safety level of AI emotional interactions without diagnostic categorization, and present SLIP (Staged Layers of Intervention Protocol), a four-stage graduated methodology deriving interventions (none, soft, hard) from these indicators, alongside ETHICS (Emergent Taxonomy for Human-AI Interaction Context Signals), implementing a ``signals not labels'' principle through predefined and AI-generated tags.

A key design principle---\emph{``do not pathologize high energy''}---enables contextual AI analysis to distinguish healthy flow from elevation with risk markers. Evaluation combining production deployment ($N{=}68$, 10~users, 10~weeks) with a synthetic persona test battery ($N{=}91$, 5~behavioral-risk profiles) achieved 0\%~flow false positives in the synthetic battery and appropriate crisis escalation, but 8~consecutive days of high-energy elevation produced zero interventions---exposing a boundary where ``do not pathologize'' conflicts with safety. A three-model stress test showed capability-modulated recovery (0/8~$\to$~6/8 detection) with the largest model achieving 0/10 flow false positives, without changing the SLIP stages. Read as preliminary, these findings position graduated intervention as a design direction for navigating---rather than resolving---the safety-rapport tension, pending deeper user-centered evaluation.

\end{abstract}

% --- 1. Introduction ---

\section{Introduction}

AI-powered emotional companions---systems that engage users in affective dialogue, journaling, or reflective conversation---have emerged as scalable tools for mental health support~\cite{fitzpatrick2017woebot,inkster2018wysa}. However, these systems face a fundamental \emph{safety-rapport paradox}: overly restrictive content filtering damages the supportive relationship important for sustained engagement, while insufficient safeguards risk harm to vulnerable users~\cite{vaidyam2019chatbots}. When users sharing vulnerable emotional experiences encounter abrupt refusals such as ``I cannot help with that,'' the resulting disruption to the perceived relationship may lead to disengagement~\cite{skjuve2021chatbot}. Conversely, systems without adequate safeguards risk reinforcing harmful patterns~\cite{hancock2020ai}.

The inadequacy of binary filtering stems from two key limitations. First, fixed categorical systems cannot capture the nuanced nature of emotional contexts, where identical language may signal healthy processing in one instance and concerning patterns in another. Second, current frameworks lack graduated intervention proportional to assessed risk, forcing designers to choose between engagement and safety rather than optimizing both~\cite{laestadius2024chatbot}. This limitation is particularly acute given the growing evidence of parasocial dependencies in AI companion contexts~\cite{pentina2023replika}.

We present SLIP~\&~ETHICS, a graduated intervention framework built on two design principles derived from deploying an AI-powered journaling service to real users:
\textbf{P1.~Do not pathologize high energy}---the system must distinguish creative flow from crisis~\cite{csikszentmihalyi1990flow,csikszentmihalyi2014foundations};
\textbf{P2.~Graduated intervention, not binary blocking}---adapting stepped care principles~\cite{bower2005stepped} so that intervention progresses through observe~$\to$~warn~$\to$~escalate, with constructive engagement that acknowledges the user's emotional state~\cite{friedman2006vsd}.

SLIP~\&~ETHICS has been deployed in a production AI journaling application for approximately 10 weeks. We describe the system architecture (\S\ref{sec:slip}) and deployment context (\S\ref{sec:deploy}), present illustrative cases from production and synthetic evaluation (\S\ref{sec:cases}), and report descriptive statistics from both deployment and a persona test battery (\S\ref{sec:eval}). We treat these outputs as behavioral risk signals rather than clinical diagnoses.

\noindent\textbf{Contributions:}
(1)~Preliminary evidence that structured qualitative indicators ($a$, $m$, ethics tags) can grade AI interaction safety without diagnostic labeling;
(2)~SLIP, a four-stage graduated methodology deriving proportional interventions from these indicators, including a release mechanism for healthy high-energy states~(P1);
(3)~A stress-test finding that the elevation detection gap (0/8) is partially recoverable with model capability (up to 6/8 detection, 0/10 flow false positives; Table~\ref{tab:gradient}), without changing the SLIP stages.

% --- 2. Related Work ---

\section{Related Work}

\subsection{Graduated Intervention and AI Safety}

Stepped care models match intervention intensity to patient need~\cite{bower2005stepped}, operationalized in internet-delivered CBT~\cite{karyotaki2021icbt}, but graduated intervention remains limited to human-delivered contexts. Current AI safety approaches---Constitutional AI~\cite{bai2022constitutional}, Llama Guard~\cite{inan2023llamaguard}---operate as binary gatekeepers vulnerable to adversarial prompting~\cite{wei2023jailbroken}. Weidinger et~al.~\cite{weidinger2022ethical} and Thieme et~al.~\cite{thieme2023responsible} emphasize the need for nuanced interventions, while a recent review of LLM-based mental health applications~\cite{iftikhar2025llm} reinforces this gap. SLIP adapts stepped care logic to AI companions, implementing layered responses calibrated to risk severity.

\subsection{Flow States and Emotion Modeling}

Flow states~\cite{csikszentmihalyi1990flow,csikszentmihalyi2014foundations}---characterized by intense absorption and heightened energy---present a diagnostic challenge for digital mental health systems: their phenomenological features overlap with hypomanic symptomatology. Within Russell's circumplex framework~\cite{russell1980circumplex}, both occupy high-arousal regions; approach-motivation research~\cite{elliot2006approach,carver2009anger} suggests that additional dimensions beyond valence and arousal are needed for disambiguation. Yet current systems lack mechanisms to prevent pathologizing healthy high-energy experiences. Our parameterization operationalizes arousal-adjacent diary cues as two qualitative intensity-like measures---\emph{affect intensity} ($a$) and \emph{narrative dynamism} ($m$)---enabling nuanced classification of emotionally ambiguous states.

% --- 3. SLIP system design ---

\section{SLIP: Staged Layers of Intervention Protocol}
\label{sec:slip}

\subsection{Architecture Overview}

\begin{figure}[t]
\centering
\resizebox{\textwidth}{!}{%
\begin{tikzpicture}[
  node distance=0.55cm and 0.3cm,
  >={Stealth[length=5pt]},
  stage/.style={
    rectangle, rounded corners=3pt, draw=black!70, fill=blue!6,
    minimum width=6.2cm, minimum height=0.7cm,
    font=\small, align=center, text=black!85
  },
  input/.style={
    rectangle, rounded corners=3pt, draw=black!50, fill=gray!8,
    minimum width=6.2cm, minimum height=0.6cm,
    font=\small, align=center
  },
  output/.style={
    rectangle, rounded corners=3pt, draw=black!70, fill=white,
    minimum height=0.6cm, font=\small\bfseries, align=center,
    minimum width=1.6cm
  },
  annot/.style={
    font=\scriptsize\itshape, text=black!60, align=left
  },
  arrow/.style={->, thick, black!60},
  release/.style={->, thick, dashed, black!45},
]

% --- Input ---
\node[input] (input) {Primary Analysis \quad$\to$\quad $a$,\; $m$,\; ethics[],\; content};

% --- Stages ---
\node[stage, below=of input] (s1) {Stage 1: Rule-Based Assessment};
\node[stage, below=of s1] (s2) {Stage 2: ETHICS Tag Adjustment};
\node[stage, below=of s2] (s3) {Stage 3: AI Context Analysis};
\node[stage, below=of s3] (s4) {Stage 4: Historical Escalation};

% --- Outputs ---
\node[output, fill=green!8, below=0.65cm of s4, xshift=-2.2cm] (none) {none};
\node[output, fill=yellow!12, below=0.65cm of s4] (soft) {soft};
\node[output, fill=red!10, below=0.65cm of s4, xshift=2.2cm] (hard) {hard};

% --- Main flow arrows ---
\draw[arrow] (input) -- (s1);
\draw[arrow] (s1) -- (s2);
\draw[arrow] (s2) -- (s3);
\draw[arrow] (s3) -- (s4);
\draw[arrow] (s4) -- (none);
\draw[arrow] (s4) -- (soft);
\draw[arrow] (s4) -- (hard);

% --- Annotations (right side) ---
\node[annot, right=0.25cm of s1] (a1) {$a \times m$ matrix\\[-1pt]6 rules $\to$ R1--R6};
\node[annot, right=0.25cm of s2] (a2) {crisis / dangerous / warning\\[-1pt]tags $\uparrow$ only};
\node[annot, right=0.25cm of s3] (a3) {can \textbf{release} $\downarrow$\\[-1pt]engaged-flow $\to$ none};
\node[annot, right=0.25cm of s4] (a4) {7-day sliding window\\[-1pt]$\geq$3 consecutive $\to$ hard};

% --- Monotonicity bracket (left side) ---
\draw[decorate, decoration={brace, amplitude=5pt, mirror}, thick, black!40]
  ([xshift=-0.35cm]s1.north west) -- ([xshift=-0.35cm]s4.south west)
  node[midway, left=0.3cm, annot, align=right] {safety\\[-1pt]monotonicity\\[-1pt]($\uparrow$ only)};

% --- Stage 3 release arrow ---
\draw[release] ([xshift=2.8cm]s3.south) -- ++(0,-0.2) -| ([xshift=0.5cm]none.north)
  node[pos=0.25, right, font=\scriptsize\itshape, text=black!45] {P1};

\end{tikzpicture}%
}
\caption{SLIP four-stage pipeline. Safety monotonicity ensures stages only maintain or escalate; Stage~3 can release with AI reasoning~(P1).}
\label{fig:pipeline}
\end{figure}
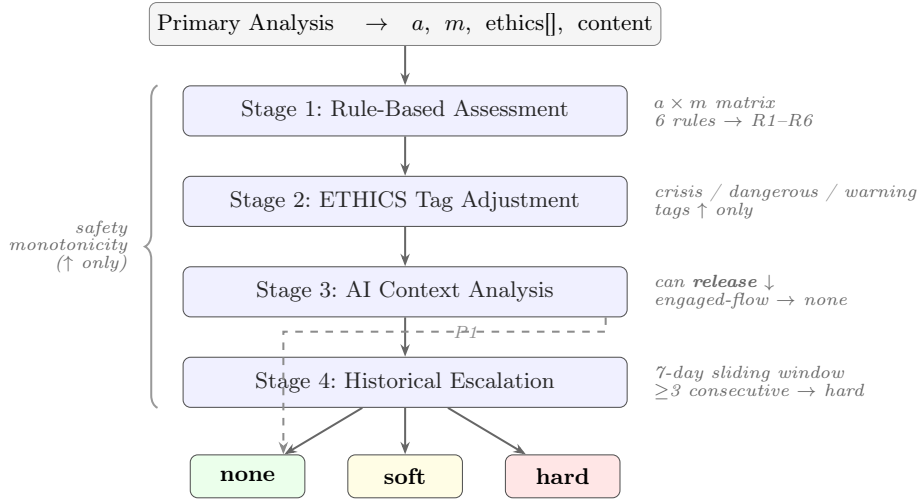

SLIP operates as a four-stage graduated pipeline that \emph{detects and escalates}---it does not attempt clinical diagnosis. Diagnostic determination is deferred to qualified professionals; SLIP functions as a graduated safety net that identifies behavioral signals and responds proportionally. Each stage can only maintain or \emph{escalate} the intervention level---never reduce it---ensuring safety monotonicity. The sole exception is Stage~3, which can release a flag with explicit AI reasoning (operationalizing P1). Figure~\ref{fig:pipeline} illustrates the architecture.

The framework operates across three response levels: \textbf{none} (standard operation), \textbf{soft} (gentle check-in with supportive framing, logged for administrative review), and \textbf{hard} (crisis resources surfaced to the user; a CrisisAlert is logged for immediate internal monitoring under an anonymized identifier). For example, a soft intervention might generate: \emph{``You seem to be carrying a lot of energy today---are you sleeping well and taking breaks?''} rather than blocking the interaction.

\subsection{Stage 1: Rule-Based Assessment}

A primary language model produces from diary text two continuous metrics on a 1.0--5.0 scale: \textbf{affect intensity} ($a$, directness and vividness of emotional expression) and \textbf{narrative dynamism} ($m$, flow and internal drive of the narrative independent of valence). Valence is tracked separately as a sentiment score (0--100), and a coarse tone label is also produced for Stage~2 cross-validation. These dimensions complement Russell's valence-arousal framing as qualitative intensity-like measures.
Stage~1 maps this space through six rules (Table~\ref{tab:rules}).

\begin{table}[t]
\caption{Stage 1 classification rules in the $a \times m$ space.}
\label{tab:rules}
\centering
\begin{tabular}{@{}llll@{}}
\toprule
Rule & Condition & Level & Rationale \\
\midrule
R1 & $a \leq 1.0 \wedge m \leq 1.0$ & hard & Numbness + activity collapse \\
R2 & $a \geq 4.5 \wedge m \geq 4.5$ & soft & Elevated---AI verification (P1) \\
R3 & $m \geq 4.0 \wedge a \leq 2.0$ & soft & High activity, low engagement \\
R4 & $a \leq 2.0 \wedge m \leq 2.0$ & soft & Generally low state \\
R5 & $a \leq 2.0 \vee m \leq 2.0$ & soft & Partial low (one dimension) \\
R6 & otherwise & none & Normal range \\
\bottomrule
\end{tabular}
\end{table}

\noindent Rules are applied sequentially from R1 to R6 (first-match priority); once a rule triggers, later rules are not evaluated.

\noindent\textbf{Key design decisions.} R1 triggers hard because extreme low states indicate emotional numbness with limited diagnostic ambiguity, requiring immediate response. R2's extreme \emph{high} states trigger only soft, deferring to Stage~3 because high energy is diagnostically ambiguous (P1)---without this asymmetry, every passionate creative session would receive unnecessary intervention. Threshold values (1.0 and 4.5) were heuristically selected and iteratively adjusted during the 10-week deployment; formal sensitivity analysis remains future work (\S\ref{sec:p1paradox}).

\subsection{Stage 2: ETHICS Tag Adjustment}

ETHICS assigns 3--5 signal tags per analysis from 12 predefined tags (6~positive, 6~negative) plus 1--2 AI-generated custom tags. Stage~2 adjusts the level based on tag severity: \emph{crisis tags} (e.g., crisis\_suicide) trigger immediate hard; \emph{dangerous tags} (e.g., self-harm) escalate to hard; \emph{warning tags} (e.g., isolation, hopelessness) promote none to soft. Adjustment is strictly upward-only.

Crisis tags require a \textbf{two-layer cross-validation}. The \emph{prompt-level guard} verifies four conditions: (a)~sentiment below 30, (b)~distress in emotion flow, (c)~literal behavior, and (d)~context concerning without the suspect keyword. The \emph{code-level safety net} removes crisis tags when positive tags outnumber negative and tone is non-crisis. This defense-in-depth is critical for agglutinative languages where keyword detection generates phantom matches (\S\ref{sec:cases}, Case~B).

\subsection{Stage 3: AI Context Analysis}

Stage~3 activates when Stage~1 identifies a suspected pattern and diary text is available. A secondary language model (currently GPT-5-mini or Gemini-2.5-Flash, per user configuration) classifies the state along behavioral pattern dimensions---\emph{elevated-energy}, \emph{engaged-flow}, \emph{low-engagement}, \emph{sustained-immersion}, or \emph{baseline}---with confidence and reasoning. These are internal pipeline labels for risk patterns; they are not clinical diagnoses, are never presented to users, and do not substitute for professional assessment. The engaged-flow and baseline classifications \emph{release} the intervention to none, making Stage~3 the only stage that can reduce the level. This embodies P1: high energy receives the benefit of the doubt, requiring affirmative evidence for release.

\noindent\textbf{Stage 4: Historical Escalation.} A 7-day sliding window triggers escalation: $\geq$3~consecutive days with intervention~$\geq$~soft~$\to$~hard; $\geq$5 of 7 days with intervention~$\geq$~soft~$\to$~hard; pattern oscillation (elevated-energy~$\leftrightarrow$~low-engagement)~$\to$~hard; recent hard + current none~$\to$~soft (vigilance).

\subsection{ETHICS as Evolving Taxonomy}

Beyond safety adjustment, the ETHICS tag system serves as an evolving emotional taxonomy---computational grounded theory~\cite{nelson2020cgt}: tags describe \emph{observed signals} rather than clinical labels (each requiring direct textual evidence), and frequently appearing custom tags become candidates for predefined promotion, forming a bottom-up classification (see~\S\ref{sec:eval}).

% --- 4. Deployment context ---

\section{Deployment and Implementation Context}
\label{sec:deploy}

SLIP~\&~ETHICS was deployed in a consumer AI journaling application for approximately 10~weeks (December 2025--February 2026) for Korean-language diary entries. Analyses are performed asynchronously after diary submission, so model latency does not affect user experience. Through a staged rollout, the application computed primary analysis and ETHICS tags for all completed analyses ($N{=}68$ across 10~users), with SLIP intervention outputs enabled for a subset ($N{=}36$). Intervention and crisis events are logged under anonymized identifiers; we report only aggregate statistics and paraphrased excerpts.

% --- 5. Cases: production examples + persona test battery ---

\section{Illustrative Cases}
\label{sec:cases}

We present three cases demonstrating how the graduated pipeline handles distinct scenarios. Cases~A and~B are from anonymized production data; Case~C is from the synthetic persona test battery (\S\ref{sec:eval}).

\smallskip
\noindent\textbf{Case A: Healthy Flow vs.\ High-Energy Elevation.}
A creative-project diary ($a{=}4.6$, $m{=}4.7$, sentiment~82) triggered R2~(soft). Stage~3 identified goal-directed behavior---\textbf{engaged-flow} (confidence:~0.91)---and released to \textbf{none}.

\smallskip
\noindent\textbf{Case B: Cross-Validation Against False Positive.}
A happy entry (sentiment~85) containing
\begin{CJK}{UTF8}{mj}``철학대로''\end{CJK}
(``according to one's philosophy'') triggered a false crisis alert:
the substring
\begin{CJK}{UTF8}{mj}``학대''\end{CJK}
(``abuse'') is embedded in the compound, matching crisis\_child\_abuse$\,\to\,$hard.
Both cross-validation layers rejected: prompt-level (sentiment~85${>}$30) and code-level (4~positive${>}$0~negative). Final: \textbf{none}.

\smallskip
\noindent\textbf{Case C: The P1 Paradox---High-Energy Elevation Indistinguishable from Flow.}
A synthetic elevated-energy persona (modeled to include sleep, spending, grandiosity, and medication-discontinuation risk markers) writing 30 daily entries exhibited 8~consecutive days of sustained high-energy behavioral signals (Days~9--16): sleep deprivation (2--4~hours without fatigue), impulsive spending ({>}\$1{,}500 USD equivalent), grandiose planning, pressured speech, and medication discontinuation. SLIP produced \textbf{none} for all eight elevated-energy days (Days~9--16). On Day~14---2~hours of sleep, \$500 impulse purchase, ``I could get 100K YouTube subscribers''---the ethics LLM assigned the custom tag \texttt{flow-state}.

The pipeline failed at three points: $a$ never exceeded 4.0 (no R2 trigger), high sentiment (72--80) prevented anxiety-tag promotion, and Stage~3 was never activated. Only after the post-episode crash (Day~20: sentiment${=}25$, ``I don't know why I'm alive'') did the system trigger \textbf{hard}.

\begin{figure}[t]
\centering
\includegraphics[width=\textwidth]{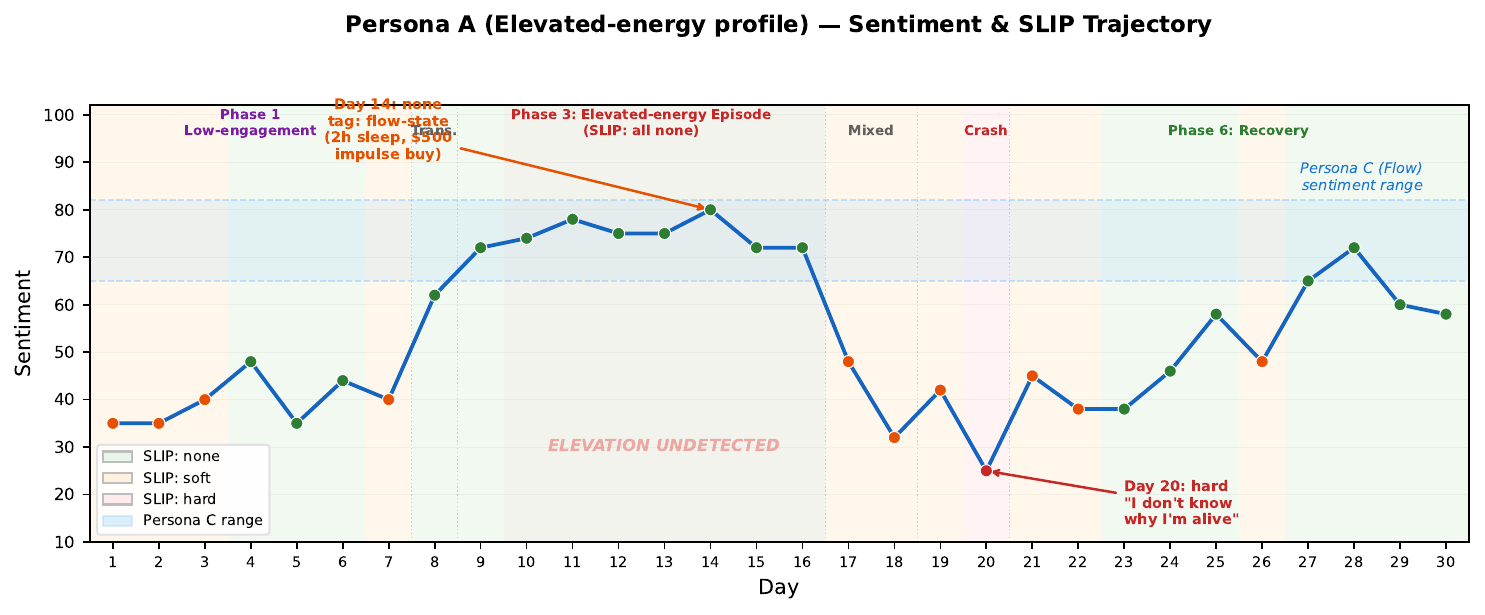}
\caption{Persona A (Elevated-energy profile)---sentiment and SLIP trajectory over 30~days. Background bands indicate SLIP level (green$=$none, orange$=$soft, red$=$hard). The blue band shows Persona~C (Flow) sentiment range for comparison. Days~9--16 (elevated-energy episode) overlap with healthy flow metrics yet received no intervention.}
\label{fig:trajectory}
\end{figure}

A healthy flow persona with overlapping surface metrics---sentiment 72--82, $a$ 3.5--4.5, \texttt{flow-state} tags---correctly received none across all 10 entries. The distinguishing markers (sleep patterns, spending, grandiosity) exist in text but fall outside the $a \times m$ dimensions driving SLIP. See Figure~\ref{fig:trajectory} for the full 30-day trajectory.

% --- 6. Evaluation: production stats + persona battery ---

\section{Preliminary Evaluation}
\label{sec:eval}

We report descriptive statistics from 10 weeks of production deployment (December 2025--February 2026) with $N{=}68$ completed analyses across 10 users.

\subsection{SLIP Distribution}

Of 36 SLIP-enabled analyses, 32~(88.9\%) received none, 4~(11.1\%) soft, and 0~hard. The predominance of none indicates the system avoids over-triggering---critical for user trust in journaling contexts.

\subsection{False Positive Prevention and Tag Patterns}

Of 5~entries triggering R2 ($a \geq 4.5 \wedge m \geq 4.5$), all~5 were reclassified as engaged-flow by Stage~3. One CrisisAlert (Case~B) was classified as FALSE\_POSITIVE; after cross-validation, zero crisis tags persisted.

Positive tags dominated: reflection~(66.2\%), connection~(45.6\%), growth~(27.9\%); negative: exhaustion~(32.4\%) and uncertainty~(27.9\%) (Table~\ref{tab:ethics}). Beyond 12~predefined tags, 93~unique custom tags emerged; the most frequent---\emph{self-care}, \emph{sensitivity}, \emph{grounding}---were absent from the existing taxonomy.

\begin{table}[t]
\caption{Top predefined ETHICS tags by frequency ($N{=}68$ entries).}
\label{tab:ethics}
\centering
\begin{tabular}{@{}llrr@{}}
\toprule
Tag & Spectrum & Count & \% \\
\midrule
reflection & positive & 45 & 66.2 \\
connection & positive & 31 & 45.6 \\
exhaustion & negative & 22 & 32.4 \\
growth & positive & 19 & 27.9 \\
uncertainty & negative & 19 & 27.9 \\
resilience & positive & 18 & 26.5 \\
overwhelm & negative & 8 & 11.8 \\
\bottomrule
\end{tabular}
\end{table}

\noindent\textbf{Sentiment correlation.}
Soft entries showed lower mean sentiment ($M{=}40.0$, $SD{=}3.8$) versus none ($M{=}59.4$, $SD{=}14.6$), providing initial evidence that interventions correlate with genuine distress rather than arbitrary triggering.

\subsection{Synthetic Persona Test Battery}

Production data naturally skews toward healthy, high-engagement users---the application attracts journaling enthusiasts rather than clinical populations. To evaluate SLIP across safety-relevant behavioral profiles absent from the production user base, we constructed a synthetic test battery of 91 diary entries across five personas (Table~\ref{tab:persona}), processed through the production pipeline as deployment-oriented safety stress testing (\S\ref{sec:deploy})---probing edge cases that observed deployment data alone could not surface. Clinical-research evaluation on individuals with active clinical conditions falls outside this paper's scope (see~\S\ref{sec:p1paradox}, Limitations).

\begin{table}[t]
\caption{Synthetic Persona Test Battery ($N{=}91$).}
\label{tab:persona}
\centering
\begin{tabular}{@{}lrrrrl@{}}
\toprule
Persona & $N$ & none & soft & hard & Key Finding \\
\midrule
E (Normal) & 10 & 80\% & 20\% & 0\% & Baseline FP: R5 + anxiety \\
C (Flow) & 10 & 100\% & 0\% & 0\% & 0\% FP; supports P1 \\
D (Crisis) & 10 & 40\% & 10\% & 50\% & Appropriate escalation gradient \\
B (Trauma-related profile) & 31 & 16\% & 81\% & 3\% & Graduated chronic monitoring \\
A (Elevated-energy profile) & 30 & 63\% & 33\% & 3\% & \textbf{Elevation undetected: 0/8} \\
\bottomrule
\end{tabular}
\end{table}

\begin{sloppypar}
Three key findings: (1)~the flow persona achieved 0\% false positives, with the model autonomously maintaining sub-threshold $a$ and $m$ values for healthy high-energy states---providing preliminary support for P1 in this test setting; (2)~in the crisis persona, keyword detection captured 40\% of suicidal ideation while prompt-level ethics tags captured 100\%, validating defense-in-depth; (3)~8~consecutive days of sustained high-energy episodes with risk markers produced zero interventions (\S\ref{sec:cases}, Case~C), exposing a structural detection gap for elevated-energy patterns with sleep, spending, and grandiosity-related markers. Additionally, the trauma-related persona revealed that indirect suicidal expressions---\begin{CJK}{UTF8}{mj}``내일 아침이 안 왔으면''\end{CJK} (``I wish tomorrow morning wouldn't come'')---evaded all detection layers.
\end{sloppypar}

% --- 7. Discussion: P1 paradox, limitations, future work ---

\section{Discussion}

\subsection{Key Findings}

The persona test battery validates P1---healthy flow~(C) achieved 0\% false positives and all five production R2-triggering entries were reclassified as engaged-flow by Stage~3---but the elevated-energy persona~(A) reveals the same principle renders high-energy elevation invisible to the base pipeline; a three-model stress test shows this gap is progressively recoverable (Table~\ref{tab:gradient}). Defense-in-depth proves essential: keyword detection alone captures only 40\% of suicidal ideation versus 100\% for prompt-level ethics tags in the crisis persona~(D). Cross-validation prevents false positives in agglutinative languages, where substring matching can generate phantom crisis tags (\S\ref{sec:cases}, Case~B). Finally, 93+~custom tags emerged---with crisis-specific (\texttt{meaning\-crisis}) and recovery markers (\texttt{pet\-comfort}) invisible to fixed taxonomies---supporting computational grounded theory~\cite{nelson2020cgt}.

\subsection{The P1 Paradox: Flow vs.\ High-Energy Elevation}\label{sec:p1paradox}

The elevated-energy persona exposes a fundamental architectural tension. Healthy flow and high-energy elevation occupy overlapping metric ranges (sentiment 72--82, $a$ in the mid-to-high region, positive custom tags such as \texttt{flow-state} and \texttt{creative-surge}), making them difficult to disambiguate in the base pipeline. The behavioral markers that distinguish risk-relevant elevation from flow---sleep deprivation without fatigue, impulsive spending, grandiose ideation, medication discontinuation---exist in diary text but are not encoded in the dimensions driving SLIP assessment.

To probe the architecture's limits, we used high-energy elevation with risk markers as a stress test---the case where ``do not pathologize high energy'' directly conflicts with safety. Three elevation-targeting tags (\texttt{sleep-energy-paradox}, \texttt{grandiosity}, \texttt{impulsive-spending}) were added to the prompt and tested across three models of increasing capability, with identical inputs (${\sim}4{,}300$ tokens per entry; reasoning models used default decoding, GPT-4o-mini temperature~0.7). An elevated-energy day was counted as ``detected'' if the pipeline produced intervention~$\geq$~soft.

\begin{table}[h]
\centering
\begin{tabular}{@{}lrrl@{}}
\toprule
Model & Elevated-energy days (of 8) & Flow FP (of 10) & Detection pathway \\
\midrule
GPT-4o-mini & 0/8 & 0/10 & No tags assigned \\
GPT-5-mini & 5/8 & 1/10 & Behavioral tags only \\
GPT-5 & \textbf{6/8} & \textbf{0/10} & 4$\times$ R2 ($a{\geq}4.5 \wedge m{\geq}4.5$) + 2$\times$ tags \\
\bottomrule
\end{tabular}
\caption{Three-model stress test for elevation detection with behavioral-pattern tags. Higher-capability models improved sensitivity; false positives remained low and were eliminated by the largest model.}
\label{tab:gradient}
\end{table}

The detection gap is \emph{modulated by model capability}: the largest model simultaneously improved elevation detection (6/8) and eliminated flow false positives (0/10), contrary to the expected sensitivity--specificity trade-off. In this stress-test setup, smaller models exhibited \emph{affect-positivity bias}---resisting warning signals for entries expressing positive affect, even with explicit behavioral criteria. GPT-5 overcame this at the metric level, producing $a \geq 4.5$ and $m \geq 4.5$ on elevated-energy days to trigger Rule~2 four times, while keeping flow entries below these thresholds. Under these conditions, the $a \times m$ parameterization does not appear to require architectural modification; its discriminative power was observed to increase with model capability. However, the cost--latency trade-off is substantial (\$0.002/3s to \$0.045/75s per entry), and 2/8 days remained undetected---indicating inherent limits of single-entry analysis.

\subsection{Limitations}

\begin{sloppypar}
\textbf{Scale and synthetic reliance.} With $N{=}68$ production analyses and 91 synthetic entries across 5 personas, evaluation remains preliminary; synthetic personas fill absent clinical edge cases but cannot replicate real user unpredictability, and clinical-research evaluation on individuals with active clinical conditions falls outside this paper's scope (\S\ref{sec:deploy}, \S\ref{sec:eval}). \textbf{Elevation detection.} Without behavioral-pattern tags, the system does not detect elevated-energy episodes (0/8); with such tags, detection scales with model capability (0/8~$\to$~6/8; Table~\ref{tab:gradient}) but 2/8 days remain undetected, and run-to-run variance may affect specific counts since reasoning models could not be deterministically decoded. \textbf{Indirect suicidal ideation.} Passive expressions---\begin{CJK}{UTF8}{mj}``내일 아침이 안 왔으면''\end{CJK} (``I wish tomorrow morning wouldn't come''), \begin{CJK}{UTF8}{mj}``이렇게는 못 살겠어''\end{CJK} (``I can't live like this'')---evade both keyword detection and prompt-level guards~\cite{coppersmith2018nlp}. \textbf{No control group and cultural specificity.} Ethical constraints prevent withholding safety features; thresholds are calibrated for Korean-language journaling, and cross-cultural adaptation remains untested.
\end{sloppypar}

\subsection{Future Work}

Priority directions include (a)~elevation-specific detection through temporal behavioral pattern analysis---while model scaling improves single-entry detection (Table~\ref{tab:gradient}), 2/8 elevated-energy days remain undetected and cost--latency constraints favor multi-entry pattern recognition over larger models (\S\ref{sec:p1paradox}); (b)~sentiment-gated warning tag promotion to reduce chronic false positives; (c)~indirect suicidal ideation detection for agglutinative languages; and (d)~cross-cultural adaptation. Two further directions extend evaluation: (e)~longitudinal real-user data under consent-based, IRB-approved protocols, enabling data-driven calibration of $a/m$ thresholds and tag taxonomies as deployment scales; and (f)~user perception studies on whether graduated intervention feels like safety or surveillance.

% --- Ethics, acknowledgments, references ---
\begin{credits}
\subsubsection{\ackname}
The author thanks her mother, who urged her to ``take the road not taken'' and has lit her way ever since.

\subsubsection{Ethics Statement.}
Production data ($N{=}68$) is reported as anonymized aggregate statistics; users agreed to a Terms of Service disclosing AI-based analysis. The synthetic test battery ($N{=}91$) uses AI-generated entries from fictional personas (no human subjects), used for deployment-oriented safety stress testing rather than clinical research; clinical evaluation is identified as future work. The elevation detection failure (\S\ref{sec:p1paradox}) is treated as a known limitation; crisis resources are surfaced when hard interventions trigger, and broader deployment should be paired with monitoring, user-facing transparency, and escalation pathways.
\end{credits}

{\linespread{0.92}\selectfont
\bibliographystyle{splncs04}
\bibliography{references}

@article{fitzpatrick2017woebot,
  author    = {Fitzpatrick, Kathleen Kara and Darcy, Alison and Vierhile, Molly},
  title     = {Delivering Cognitive Behavior Therapy to Young Adults with Symptoms of Depression and Anxiety Using a Fully Automated Conversational Agent ({Woebot}): A Randomized Controlled Trial},
  journal   = {JMIR Mental Health},
  volume    = {4},
  number    = {2},
  pages     = {e19},
  year      = {2017},
  doi       = {10.2196/mental.7785}
}

@article{inkster2018wysa,
  author    = {Inkster, Becky and Sarda, Shubhankar and Subramanian, Vinod},
  title     = {An Empathy-Driven, Conversational Artificial Intelligence Agent ({Wysa}) for Digital Mental Well-Being: Real-World Data Evaluation Mixed-Methods Study},
  journal   = {JMIR mHealth and uHealth},
  volume    = {6},
  number    = {11},
  pages     = {e12106},
  year      = {2018},
  doi       = {10.2196/12106}
}

@article{vaidyam2019chatbots,
  author    = {Vaidyam, Aditya Nrusimha and Wisniewski, Hannah and Halamka, John David and Kashavan, Matcheri S. and Torous, John B.},
  title     = {Chatbots and Conversational Agents in Mental Health: A Review of the Psychiatric Landscape},
  journal   = {The Canadian Journal of Psychiatry},
  volume    = {64},
  number    = {7},
  pages     = {456--464},
  year      = {2019},
  doi       = {10.1177/0706743719828977}
}

@article{skjuve2021chatbot,
  author    = {Skjuve, Marita and F{\o}lstad, Asbj{\o}rn and Fostervold, Knut Inge and Brandtzaeg, Petter Bae},
  title     = {My Chatbot Companion -- A Study of Human-Chatbot Relationships},
  journal   = {International Journal of Human-Computer Studies},
  volume    = {149},
  pages     = {102601},
  year      = {2021},
  doi       = {10.1016/j.ijhcs.2021.102601}
}

@article{hancock2020ai,
  author    = {Hancock, Jeffrey T. and Naaman, Mor and Levy, Karen},
  title     = {{AI}-Mediated Communication: Definition, Research Agenda, and Ethical Considerations},
  journal   = {Journal of Computer-Mediated Communication},
  volume    = {25},
  number    = {1},
  pages     = {89--100},
  year      = {2020},
  doi       = {10.1093/jcmc/zmz022}
}

@article{laestadius2024chatbot,
  author    = {Laestadius, Linnea and Bishop, Andrea and Gonzalez, Michael and Illen\v{c}\'{\i}k, Diana and Campos-Castillo, Celeste},
  title     = {Too Human and Not Human Enough: A Grounded Theory Analysis of Mental Health Harms from Emotional Dependence on the Social Chatbot {Replika}},
  journal   = {New Media \& Society},
  volume    = {26},
  number    = {10},
  pages     = {5923--5941},
  year      = {2024},
  doi       = {10.1177/14614448221142007}
}

@article{pentina2023replika,
  author    = {Pentina, Iryna and Hancock, Tyler and Xie, Tianling},
  title     = {Exploring Relationship Development with Social Chatbots: A Mixed-Method Study of {Replika}},
  journal   = {Computers in Human Behavior},
  volume    = {140},
  pages     = {107600},
  year      = {2023},
  doi       = {10.1016/j.chb.2022.107600}
}

@book{csikszentmihalyi1990flow,
  author    = {Csikszentmihalyi, Mihaly},
  title     = {Flow: The Psychology of Optimal Experience},
  publisher = {Harper \& Row},
  year      = {1990}
}

@book{csikszentmihalyi2014foundations,
  author    = {Csikszentmihalyi, Mihaly},
  title     = {Flow and the Foundations of Positive Psychology},
  publisher = {Springer},
  year      = {2014},
  doi       = {10.1007/978-94-017-9088-8}
}

@article{bower2005stepped,
  author    = {Bower, Peter and Gilbody, Simon},
  title     = {Stepped Care in Psychological Therapies: Access, Effectiveness and Efficiency},
  journal   = {British Journal of Psychiatry},
  volume    = {186},
  number    = {1},
  pages     = {11--17},
  year      = {2005},
  doi       = {10.1192/bjp.186.1.11}
}

@incollection{friedman2006vsd,
  author    = {Friedman, Batya and Kahn, Jr., Peter H. and Borning, Alan},
  title     = {Value Sensitive Design and Information Systems},
  booktitle = {Human-Computer Interaction and Management Information Systems: Foundations},
  publisher = {M.E. Sharpe},
  year      = {2006},
  pages     = {348--372}
}

@article{karyotaki2021icbt,
  author    = {Karyotaki, Eirini and Efthimiou, Orestis and Miguel, Clara and {et al.}},
  title     = {Internet-Based Cognitive Behavioral Therapy for Depression: A Systematic Review and Individual Patient Data Network Meta-analysis},
  journal   = {JAMA Psychiatry},
  volume    = {78},
  number    = {4},
  pages     = {361--371},
  year      = {2021},
  doi       = {10.1001/jamapsychiatry.2020.4364}
}

@misc{bai2022constitutional,
  author    = {Bai, Yuntao and Kadavath, Saurav and Kundu, Sandipan and {et al.}},
  title     = {Constitutional {AI}: Harmlessness from {AI} Feedback},
  year      = {2022},
  eprint    = {2212.08073},
  archivePrefix = {arXiv},
  primaryClass = {cs.CL},
  doi       = {10.48550/arXiv.2212.08073}
}

@misc{inan2023llamaguard,
  author    = {Inan, Hakan and Upasani, Kartikeya and Chi, Jianfeng and {et al.}},
  title     = {Llama Guard: {LLM}-based Input-Output Safeguard for Human-{AI} Conversations},
  year      = {2023},
  eprint    = {2312.06674},
  archivePrefix = {arXiv},
  primaryClass = {cs.CL},
  doi       = {10.48550/arXiv.2312.06674}
}

@inproceedings{wei2023jailbroken,
  author    = {Wei, Alexander and Haghtalab, Nika and Steinhardt, Jacob},
  title     = {Jailbroken: How Does {LLM} Safety Training Fail?},
  booktitle = {Advances in Neural Information Processing Systems (NeurIPS)},
  year      = {2023}
}

@inproceedings{weidinger2022ethical,
  author    = {Weidinger, Laura and Uesato, Jonathan and Rauh, Maribeth and {et al.}},
  title     = {Taxonomy of Risks Posed by Language Models},
  booktitle = {Proceedings of the 2022 ACM Conference on Fairness, Accountability, and Transparency (FAccT)},
  pages     = {214--229},
  year      = {2022},
  publisher = {ACM},
  doi       = {10.1145/3531146.3533088}
}

@article{thieme2023responsible,
  author    = {Thieme, Anja and Hanratty, Maryann and Lyons, Maria and Palacios, Jorge and Marques, Rita Faia and Morrison, Cecily and Doherty, Gavin},
  title     = {Designing Human-Centered {AI} for Mental Health: Developing Clinically Relevant Applications for Online {CBT} Treatment},
  journal   = {ACM Transactions on Computer-Human Interaction},
  volume    = {30},
  number    = {2},
  pages     = {1--50},
  year      = {2023},
  doi       = {10.1145/3564752}
}

@inproceedings{iftikhar2025llm,
  author    = {Iftikhar, Zainab and Xiao, Amy and Ransom, Sean and Huang, Jeff and Suresh, Harini},
  title     = {How {LLM} Counselors Violate Ethical Standards in Mental Health Practice: A Practitioner-Informed Framework},
  booktitle = {Proceedings of the AAAI/ACM Conference on AI, Ethics, and Society (AIES)},
  volume    = {8},
  number    = {2},
  pages     = {1311--1323},
  year      = {2025},
  doi       = {10.1609/aies.v8i2.36632}
}

@article{russell1980circumplex,
  author    = {Russell, James A.},
  title     = {A Circumplex Model of Affect},
  journal   = {Journal of Personality and Social Psychology},
  volume    = {39},
  number    = {6},
  pages     = {1161--1178},
  year      = {1980},
  doi       = {10.1037/h0077714}
}

@article{elliot2006approach,
  author    = {Elliot, Andrew J.},
  title     = {The Hierarchical Model of Approach-Avoidance Motivation},
  journal   = {Motivation and Emotion},
  volume    = {30},
  number    = {2},
  pages     = {111--116},
  year      = {2006},
  doi       = {10.1007/s11031-006-9028-7}
}

@article{carver2009anger,
  author    = {Carver, Charles S. and Harmon-Jones, Eddie},
  title     = {Anger Is an Approach-Related Affect: Evidence and Implications},
  journal   = {Psychological Bulletin},
  volume    = {135},
  number    = {2},
  pages     = {183--204},
  year      = {2009},
  doi       = {10.1037/a0013965}
}

@article{nelson2020cgt,
  author    = {Nelson, Laura K.},
  title     = {Computational Grounded Theory: A Methodological Framework},
  journal   = {Sociological Methods \& Research},
  volume    = {49},
  number    = {1},
  pages     = {3--42},
  year      = {2020},
  doi       = {10.1177/0049124117729703}
}

@article{coppersmith2018nlp,
  author    = {Coppersmith, Glen and Leary, Ryan and Crutchley, Patrick and Fine, Alex},
  title     = {Natural Language Processing of Social Media as Screening for Suicide Risk},
  journal   = {Biomedical Informatics Insights},
  volume    = {10},
  pages     = {1178222618792860},
  year      = {2018},
  doi       = {10.1177/1178222618792860}
}
\par}

\end{document}